\documentclass[10pt,aps,pre,onecolumn,notitlepage,superscriptaddress,showpacs,preprintnumbers]{revtex4-1}

\usepackage{amsmath,amssymb,amsfonts}
\usepackage{graphicx,color}
\usepackage{verbatim}
\usepackage{float}
\usepackage{dcolumn}
\usepackage{natbib}
\usepackage{bm}
\usepackage{hyperref}

\newcommand{\ie}{\emph{i.e.}}
\newcommand{\eg}{\emph{e.g.}}
\newcommand{\ea}{\emph{et al.}}
\newcommand{\avg}[1]{\langle #1\rangle}

\begin{document}

\title{Statistical validation of financial time series via visibility graph}

\author{Matteo Serafino}
\affiliation{Dipartimento di Fisica, Universit\`a ``Sapienza'', Piazzale Aldo Moro 5, 00185 Rome - Italy}
\author{Andrea Gabrielli}
\affiliation{Istituto dei Sistemi Complessi (ISC)-CNR UoS Universit\`a ``Sapienza'', Piazzale Aldo Moro 5, 00185 Rome - Italy}
\affiliation{IMT School for Advanced Studies, Piazza S.Francesco 19, 55100 Lucca - Italy}
\affiliation{London Institute for Mathematical Sciences, 35a South Street, W1K 2XF London - United Kingdom}
\author{Guido Caldarelli}
\affiliation{IMT School for Advanced Studies, Piazza S.Francesco 19, 55100 Lucca - Italy}
\affiliation{Istituto dei Sistemi Complessi (ISC)-CNR UoS Universit\`a ``Sapienza'', Piazzale Aldo Moro 5, 00185 Rome - Italy}
\affiliation{London Institute for Mathematical Sciences, 35a South Street, W1K 2XF London - United Kingdom}
\affiliation{European Centre for Living Technology (ECLT), San Marco 2940, 30124 Venice - Italy}
\author{Giulio Cimini}\email{giulio.cimini@imtlucca.it}
\affiliation{IMT School for Advanced Studies, Piazza S.Francesco 19, 55100 Lucca - Italy}
\affiliation{Istituto dei Sistemi Complessi (ISC)-CNR UoS Universit\`a ``Sapienza'', Piazzale Aldo Moro 5, 00185 Rome - Italy}

\begin{abstract}
Statistical physics of complex systems exploits network theory not only to model, but also to effectively extract information from many dynamical real-world systems. 
A pivotal case of study is given by financial systems: market prediction represents an unsolved scientific challenge yet with crucial implications for society, 
as financial crises have devastating effects on real economies. Thus, nowadays the quest for a robust estimator of market efficiency is both a scientific and institutional priority. 
In this work we study the visibility graphs built from the time series of several trade market indices. 
We propose a validation procedure for each link of these graphs against a null hypothesis derived from ARCH-type modeling of such series. 
Building on this framework, we devise a market indicator that turns out to be highly correlated and even predictive of financial instability periods.
\end{abstract}

\maketitle

\section{Introduction}

Network theory represents an effective tool for the description of complex systems \cite{caldarelli2007scale-free}. 
In particular, the use of appropriate network ensembles allows to apply standard statistical physics to a variety of different phenomena (see, \eg, \cite{cimini2015systemic,straka2017grand}). 
In this work we focus on the important case study of financial systems. Within this field, key issues faced by researchers, asset managers and policymakers 
consist in assessing market risk and forecasting future crises, as the presence of financial instability harms economic activities and societal welfare \cite{mishkin1999global}. 
Financial time series analysis plays an important role in this context, by allowing to extract market trends and to gain insights on market dynamics \cite{tsay2005analysis}. 
Time series of market indices are typically modeled as realizations of discrete-time stochastic processes \cite{chakraborti2011econophysics}. Under the efficient market hypothesis 
that the current value of an index contains all available information on the market, time series are modeled as a martingale, leading to a Brownian motion of the index \cite{malkiel1970efficient}.
Empirical regularities (\ie, stylized facts) of financial time series \cite{cont2001empirical} are instead described by the ARCH-type family of models. 
The ARCH (autoregressive conditional heteroskedasticity) \cite{engle1982autoregressive} and GARCH (generalized ARCH) \cite{bollerslev1986generalized} models account for volatility clustering 
and fat-tail behaviors, and can be extended in different fashions---for instance, GJR-GARCH \cite{glosten1993relation} additionally accounts for leverage effects.

In order to reveal complex patterns of time series beyond commonly used models, recently several methods that map time series into networks have been developed (see \cite{gao2016complex} for an overview). 
Some of them are based on energy landscape properties \cite{doye2002network,gfeller2007complex}, but the most effective example is the {\em visibility graph} (VG) \cite{lacasa2008from,nunez2012visibility}, 
an algorithm which takes a time series as a landscape and connects every point of the series with all those that can be seen from the top of it.  
The VG has been used to describe stochastic, fractal and chaotic univariate time series \cite{luque2009horizontal,lacasa2010description}, as well as multivariate ones \cite{lacasa2015network}. 
In the context of financial series, the use of the VG already led to important insights \cite{stephen2015visibility}. For instance, Yang \ea~\cite{yang2009visibility} 
found that exchange rate series map into scale-free networks with hierarchical structure, which can be used to quantify the Hurst exponent of the series \cite{rasheed2004hurst}. 
Yan and van Serooskerken \cite{yan2015forecasting} showed that the VG connectivity can be used to measure the magnitude of the super-exponential change of stock prices and to forecast financial extremes. 
Zhang \ea~\cite{zhang2017visibility} studied VG of time series generated by autoregressive models AR(1) and AR(2) to study the relation 
between the correlation length of the series and the exponential rate of the VG connectivity distribution. Gon\c{c}alves \ea~\cite{goncalves2017quantifying} analysed VG connectivities 
from global trade market series with Information Theory concepts to extract a quantifier of market risk, which shows to be highly correlated with financial instability periods. 

In this work we propose a novel validation framework \cite{serrano2009extracting,tumminello2011statistically,orsini2015quantifying,gualdi2016statistically,saracco2017inferring} 
for VG networks built from financial time series, in order to obtain a statistically significant signal reflecting complex non-linear patterns of such series 
and thus providing a signature for financial instability. 
We achieve this by assessing each VG connection against its expected probability of occurrence under a null hypothesis derived from the ARCH-type family of models. 
These models are particularly fit for our purposes, as they account for stylized facts of financial markets but are otherwise maximally random. 
For instance, they cannot capture the super-exponential growth (drop) of market prices ending in crashes (rallies) \cite{johansen2000crashes,sornette2003critical,sornette2006predictability,yan2012diagnosis}, 
which is instead reflected by a dramatic change in the connectivity of the corresponding time series VG before such financial extremes \cite{yan2015forecasting}. 
Indeed, we show that the number of validated connections increases in correspondence of financial crisis periods for several global market indices. 
Building on this framework, we finally introduce a metrics of market (in)stability, which turns out to be a robust predictor of financial turmoil periods.

\section{Methods}

Let us consider a generic asset whose price is described by a time series $Y=\{y_t\}_{t=1}^{T}$, with $T$ denoting its length. As usually done in the financial literature, 
we assume statistical stationarity of price returns $r_t= y_{t+1}/y_t-1$ and define the historical volatility $\sigma_t$ as the standard deviation of returns up to time $t$.

\subsection{Mapping time series into networks}

Consider a generic time series $Y=\{y_t\}_{t=1}^{T}$. The series can be seen as a landscape, in which each point $t$ of the series has a height equal to its value $y_t$. 
The visibility graph (VG) algorithm \cite{lacasa2008from} associates a node $i$ to each time $t$, 
and connects two nodes $i<j$ if a straight line between $y_i$ and $y_j$ can be drawn in the landscape without intersecting any intermediate data height $y_k$ with $i<k<j$. 
More formally, the generic element of the network adjacency matrix $a_{ij}=1$ (meaning that $i$ and $j$ are connected) if
\begin{equation}\label{eq:VG}
\frac{y_k-y_i}{k-i}<\frac{y_j-y_i}{j-k}\qquad\forall (i<k<j)
\end{equation}
and $a_{ij}=0$ otherwise. Note that the VG algorithm always connects nearest neighbor nodes, \ie, $|i-j|=1\Rightarrow a_{ij}=1$ $\forall i,j$. 
Additionally, the generated network is undirected by construction. We define the degree of a node as the number of its connections, \ie, $d_i=\sum_{j\neq i} a_{ij}$.

The complement of the VG is the invisibility graph (IVG) \cite{yan2015forecasting}, obtained by inverting the inequality in eq. \eqref{eq:VG}: $\bar{a}_{ij}=1$ if
\begin{equation}\label{eq:IVG}
\frac{y_k-y_i}{k-i}>\frac{y_j-y_i}{j-k}\qquad\forall (i<k<j)
\end{equation}
and $\bar{a}_{ij}=0$ otherwise. Again, the degree of a node is $\bar{d}_i=\sum_{j\neq i} \bar{a}_{ij}$. 
Note that we use the bar symbol to denote quantities in this case, as the IVG is the {\em complement} of the VG: 
$a_{ij}+\bar{a}_{ij}=1$ $\forall i\neq j$, \ie, apart from self connections which are absent in both cases.

\subsection{Null model of time series}

In the VG two nodes are connected as soon as the visibility criterion is satisfied. 
Thus, also random fluctuations within the time series can generate network connections. 
In order to identify those connections that correspond to unstable financial periods, we need a null model for the time series 
which is maximally random but also able to reproduce the empirical regularities of financial markets. 
In this way the model will be able to capture ``typical'' fluctuations but will fail in reproducing non-linear behaviors such as a super-exponential growth of prices. 
In this work, as null model we use the GJR-GARCH \cite{glosten1993relation}, which accounts for leptokurtosis, volatility clustering and leverage effect as stylized facts. 
We remark that the framework we propose here is general and other null models can be used as well (\ie, basically any variant of ARCH). 

Assuming the stationarity of the process, the GJR-GARCH(1,1,1) model defines returns and conditional variance (\ie, volatility squared) as follows:
\begin{eqnarray}\label{eq:garch}
r_t&\equiv& a_{t}=\sigma_{t} \epsilon_{t}\\
\sigma_t^2&=&\alpha_{0}+\alpha_1 a_{t-1}^2 + \beta_1 \sigma_{t-1}^2 + \gamma_1 a_{t-1}^2 I_{t-1}
\end{eqnarray} 
where $I_{t-1}=1$ if $a_{t-1}<0$ and is zero otherwise, and the $\epsilon_{t}$ are independent and identically distributed random variables with zero mean and unitary variance 
(typically following the standard normal or standardized t-Student distribution). The model parameters $(\alpha_{0},\alpha_1,\beta_1,\gamma_1)$ 
are estimated from the empirical time series using a maximum likelihood technique \cite{johansen1990maximum,bollerslev1992quasi}.
Note that the unconditional mean and volatility of the model depend on such parameters as $E(a_t)=0$ and $E[a_t^2]=\alpha_0/(1-\alpha_1-\beta_1-\gamma_1/2)$, 
and thus the model is stationary when $\alpha_0>0$ and $\alpha_1+\beta_1+\gamma_1/2<1$ \cite{ling2002stationarity}.

Once the model is defined, we use it to generate simulated time series (we use the symbol $\sim$ to denote simulated quantities). 
In order to directly have a stationary process, we choose as initial condition the historical volatility of the empirical series $\sigma_0=\sigma_T$. 
To generate a single series, we extract $\epsilon_0$ and compute $\tilde{r}_0=\epsilon_0\sigma_{0}$; 
then, for each $t=2,\dots,T$, we generate $\epsilon_{t}$, and update $\tilde{r}_t$ and $\tilde{\sigma}_t$ as for eq. \eqref{eq:garch}. 
By construction, the simulated series has the same unconditional mean and volatility of the empirical time series.
The process is repeated $Z$ times to finally have an ensemble $\Omega$ of $Z$ time series for returns and volatilities. 

\subsection{Validation against the null model}

Once the ensemble of simulated time series is built, we use VG to map each of them into a network, and thus end up with an ensemble $\Omega$ of $Z$ networks. 
We use this ensemble to compare the VG from the empirical series with, and to build the validated network. 

Let us denote by $p_{ij}$ the ensemble average of the generic adjacency matrix element, namely the fraction of networks in $\Omega$ 
in which the connection between $i$ and $j$ appears. We argue that if this value is low, the possible presence of the corresponding connection in the empirical VG 
is not likely to be the result of random fluctuations. 
Since we do not know the probability distribution of $p_{ij}$ within $\Omega$, we use a simple percentile criterion introducing a threshold $\rho$. 
Then, whenever $a_{ij}=1$ and $p_{ij}\le\rho$, we validate the corresponding connection, otherwise we discard it. 
We repeat this process for each connection of the empirical VG to obtain the validated VG.

\section{Data}

The date considered in this work are the following global equity indices, each considered as bellwether of market and economy of the corresponding country: 
S\&P500 (USA), MERVAL (Argentina), ASE (Greece), CAC40 (France), DAX (Germany), IBEX (Spain), SMI (Switzerland), and UKX (United Kingdom).
We remark that the choice of these indices is arbitrary, and that our methodology can be applied to any financial series. 

Data for each index is collected from the Bloomberg Terminal, and covers the time span from July $11^{th}$, $1995$ to July $11^{th}$, $2017$, 
from which we removed holidays and weekends. Each element in a series represents the closing price of the index, from which we computed returns and volatility values.

\section{Results}

\subsection{Null model basics}

\begin{figure}[t]
\centering
\includegraphics[width=0.5\textwidth]{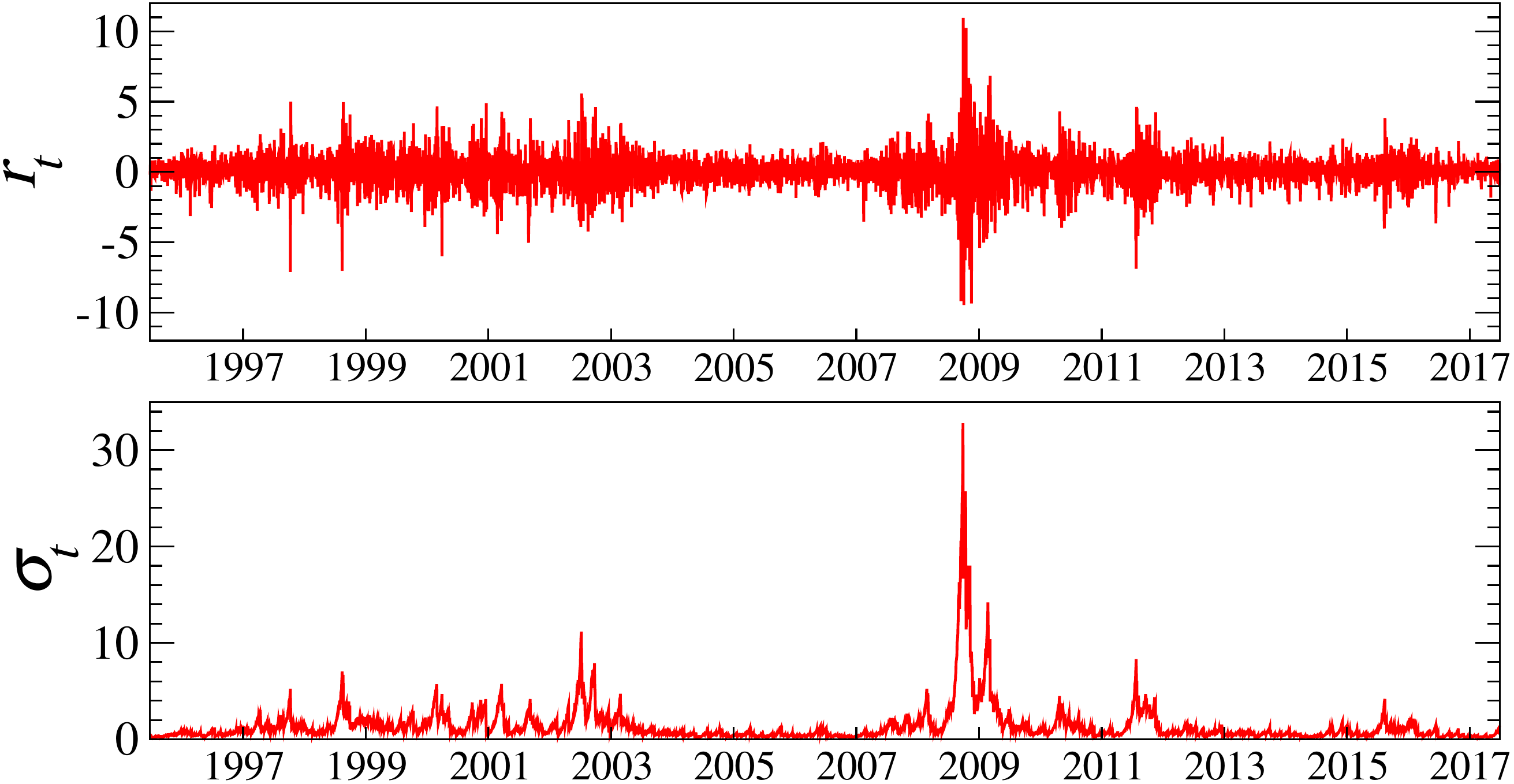}
\caption{Time series of returns (upper panel) and conditional volatility (lower panel) for S$\&$P500.}\label{fig1}
\end{figure}

In this section we discuss basic yet important properties of the null model ensemble, taking as specific example S\&P500 depicted in Figure \ref{fig1} (other indices yield qualitatively similar insights). 
We use the whole time span $T$ of data and the t-Student distribution for the white noise to build the GJR-GARCH(1,1,1) null model of the time series. 
Results for the model parameters estimated via maximum likelihood are reported in Table \ref{tab:sep}. 

\begin{table}[t]
\caption{GJR-GARCH parameters for S$\&$P500, together with values of standard error, t-statistic and degrees of freedom (dof) of the t-Student's distribution 
used in the likelihood estimation of parameters (assuming that the dimension of the sample is sufficiently large, parameters are normally distributed, so that the t-statistic test 
is equal to the parameter value over the standard error). $p$-values are computed from the $t$-statistics, showing that all parameters are statistically significant at the 5\% level 
(we indicate with $0^*$ values below the machine precision). Results for the other index are given in the appendix, Table \ref{tab:other}.}\label{tab:sep}
\begin{tabular}{c c c c c}
\hline
\textbf{parameters} & \textbf{estimate} & \textbf{std. error} & \textbf{$t$-statistic} & \textbf{$p$-value}\\ 
\hline
$\alpha_{0}$ & 0.002 & 0.001   & 2.15   & 0.0316 \\
$\alpha_{1}$ & 0     & 0       & 0     & $0^*$ \\
$\beta_{1} $ & 0.926 & 0.006   & 142.6 & $0^*$\\
$\gamma_{1}$ & 0.14  & 0.01    & 11.5  & $0^*$ \\
dof 	     &  8.9  &  1.2    & 7.8  &  $<10^{-14}$ \\
\hline\end{tabular}
\end{table}

Simulated time series are computed using these parameters and the historical volatility as initial condition, as described above. 
This approach is self-consistent, in the sense that if we use these simulated series to estimate again the model parameters, 
we obtain results that fall within the confidence interval of parameters estimated from the empirical series (\ie, those of Table \ref{tab:sep}). 
We recall that using historical volatility as initial condition allows to obtain series which are statistically stationary from the beginning. 
In fact, a different initial condition would lead to a transient period until stationarity, whose time scale is given by 
$\tau=[-\ln(\alpha_1+\beta_1+\gamma_1/2)]^{-1}$ \cite{bollerslev1986generalized} (equal to approximately 250 days for S\&P500).

In the reminder of the paper, we will focus on time series of volatility values---which are converted into networks via VG, thus using eq. \eqref{eq:VG} with $y_t=\sigma_t$. 

\begin{figure}[t]
\centering
\includegraphics[width=0.66\textwidth]{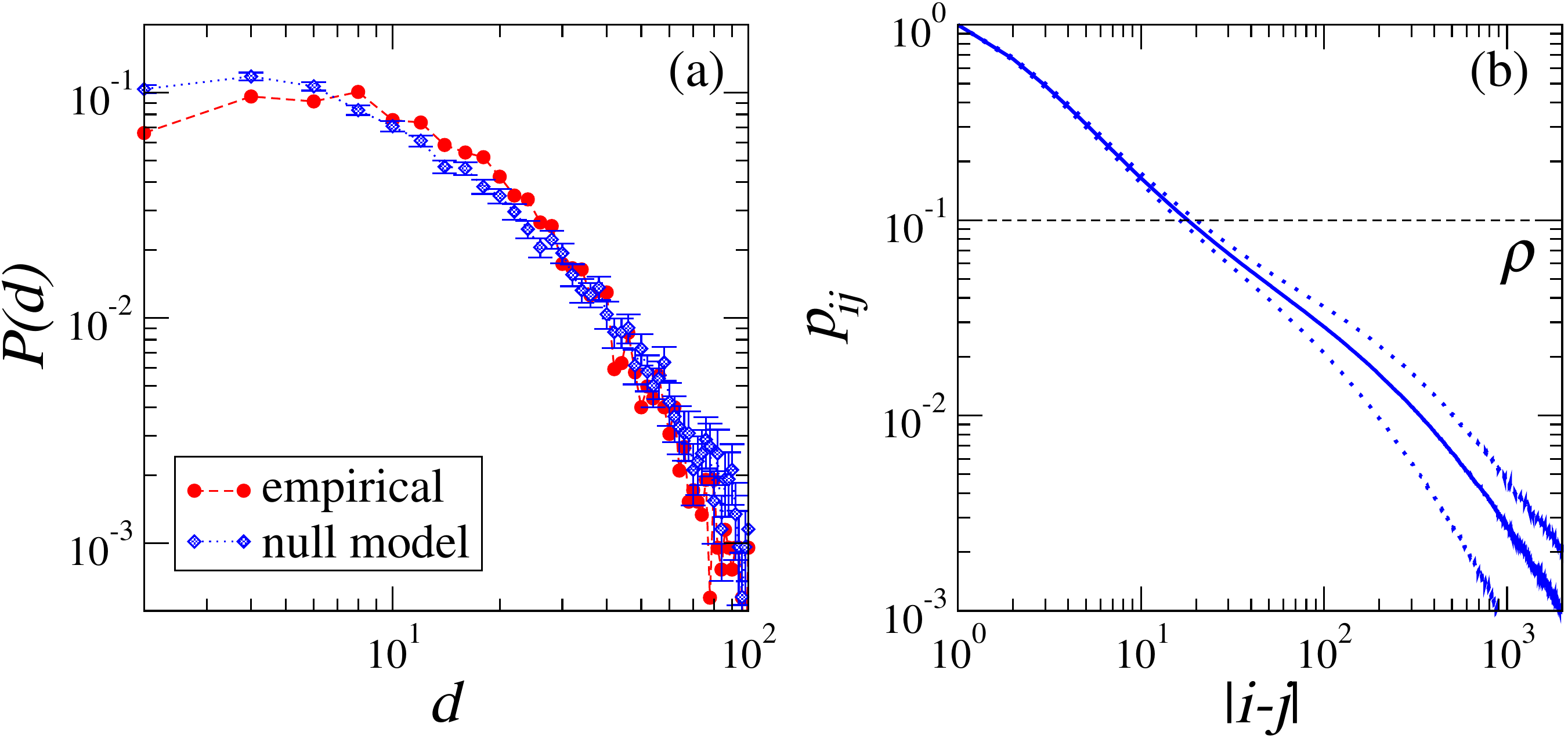}
\caption{Panel (a): degree distribution of VG from the empirical S\&P500 time series of volatilities, and from the corresponding null model ensemble.
Panel (b): connection probability $p_{ij}$ in the S\&P500 null model as a function of the absolute distance $|i-j|$ between nodes, averaged over all $(i,j)$ pairs. 
Dotted lines mark the confidence interval. The horizontal dashed line marks the validation threshold $\rho$.}\label{fig2}
\end{figure}

\paragraph*{Ensemble cardinality.} The null model ensemble should contain a number $Z$ of samples large enough to obtain a proper estimate of connection probabilities. 
In order to study the stability properties of the ensemble as a function of $Z$, we compute the {\em variation of information} \cite{meila2003comparing} between pairs of degree distributions 
averaged over ensemble realizations of different sizes. Note that since the null model that we use is maximally random, measures derived from Shannon entropy 
cannot capture differences between simulated time series: the variation of information between any two ensemble converges to zero for $T\to\infty$, 
and residual discrepancies are only due to finite-size effects. Yet variation of information values do fluctuate around such a stationary value, 
and the coefficient of variation is about 15\%, 10\% and 5\% for ensembles with $Z=10^1$, $10^2$ and $10^3$, respectively. 
To get to 2\%, $Z=10^4$ is needed. To have a good compromise between ensemble reliability and computational efficiency, in the following we set $Z=3000$.

\paragraph*{Degree distributions.} Figure \ref{fig2}(a) shows the degree distribution of the VG from the empirical S\&P500 time series of volatilities, 
compared to that from the corresponding null model ensemble. A Wilcoxon rank-sum test for the two distributions returns a $p$-value of 0.4041, 
providing no evidence that the two samples derive from different populations (the same outcome is observed for the other financial indices we consider). 
Thus, the empirical and null model degree distributions are compatible: the null model preserves the degree distribution. 
This also suggests that using a different null model scheme, obtained by reshuffling the connections in the VG but preserving the degree of each node, 
known as the {\em configuration model} \cite{newman2001random}, may represent a valid alternative. 
Note however that differences of empirical degree distributions computed on distinct (and shorter) time windows seem to provide signals of financial instability \cite{goncalves2017quantifying}.

\paragraph*{Null model connectivities.} The VG is invariant under affine transformations of the series data \cite{lacasa2008from}, in particular under horizontal translations.
At the same time, the null model we use is maximally random but for reproducing stylized facts. As such, all nodes in the null network are equivalent, 
and the connection probability between any two nodes should depend only on their distance. This is shown in Figure \ref{fig2}(b). 
Given this picture, our validation procedure assumes an intuitive meaning: roughly, we deem as statistically significant the connections in the empirical VG 
which are established between nodes at a distance greater than a critical value, determined by the threshold $\rho$ (the horizontal line in the same figure).

\subsection{Sliding window validation}

\begin{figure*}[t]
\centering
\includegraphics[width=0.48\textwidth]{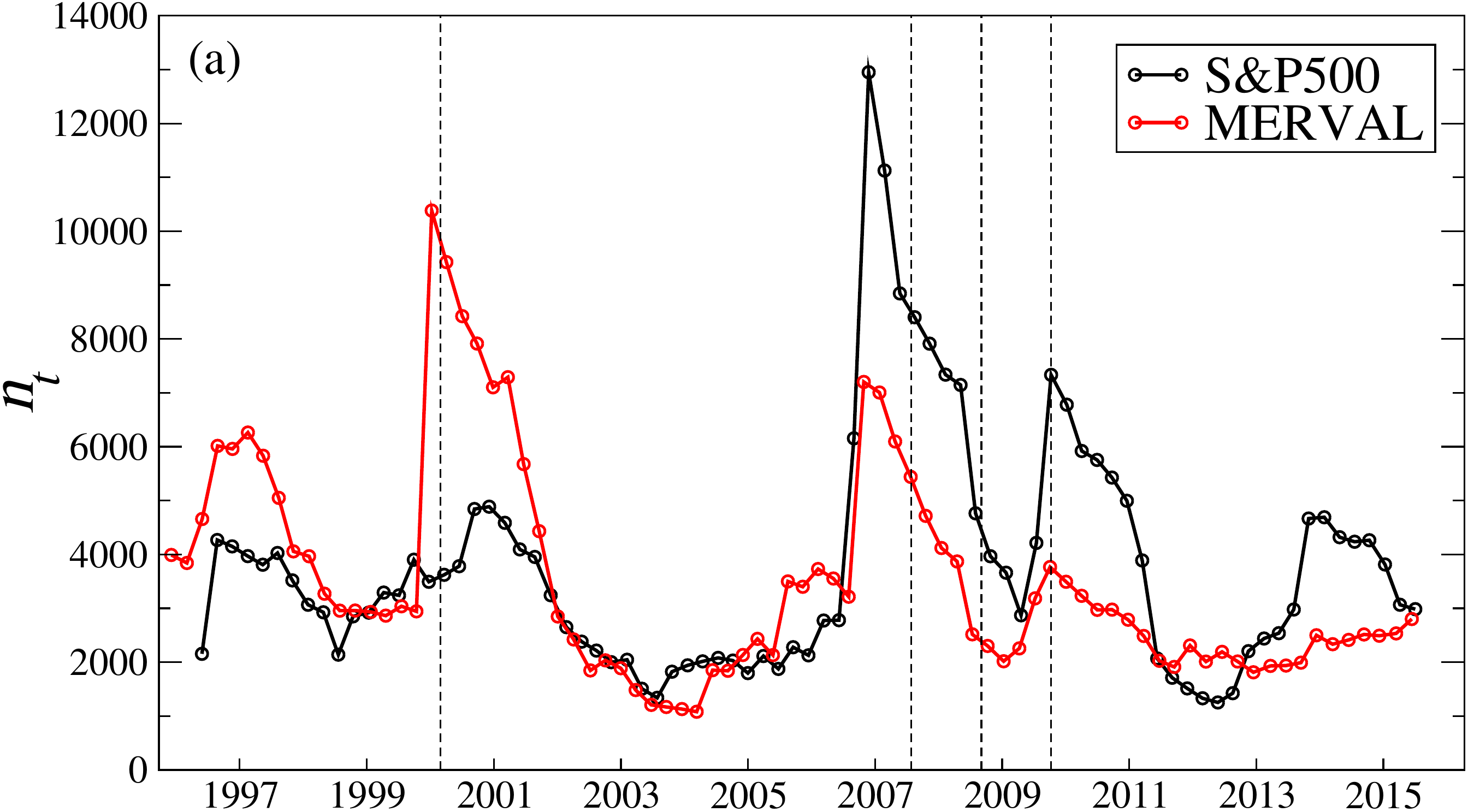}
\includegraphics[width=0.48\textwidth]{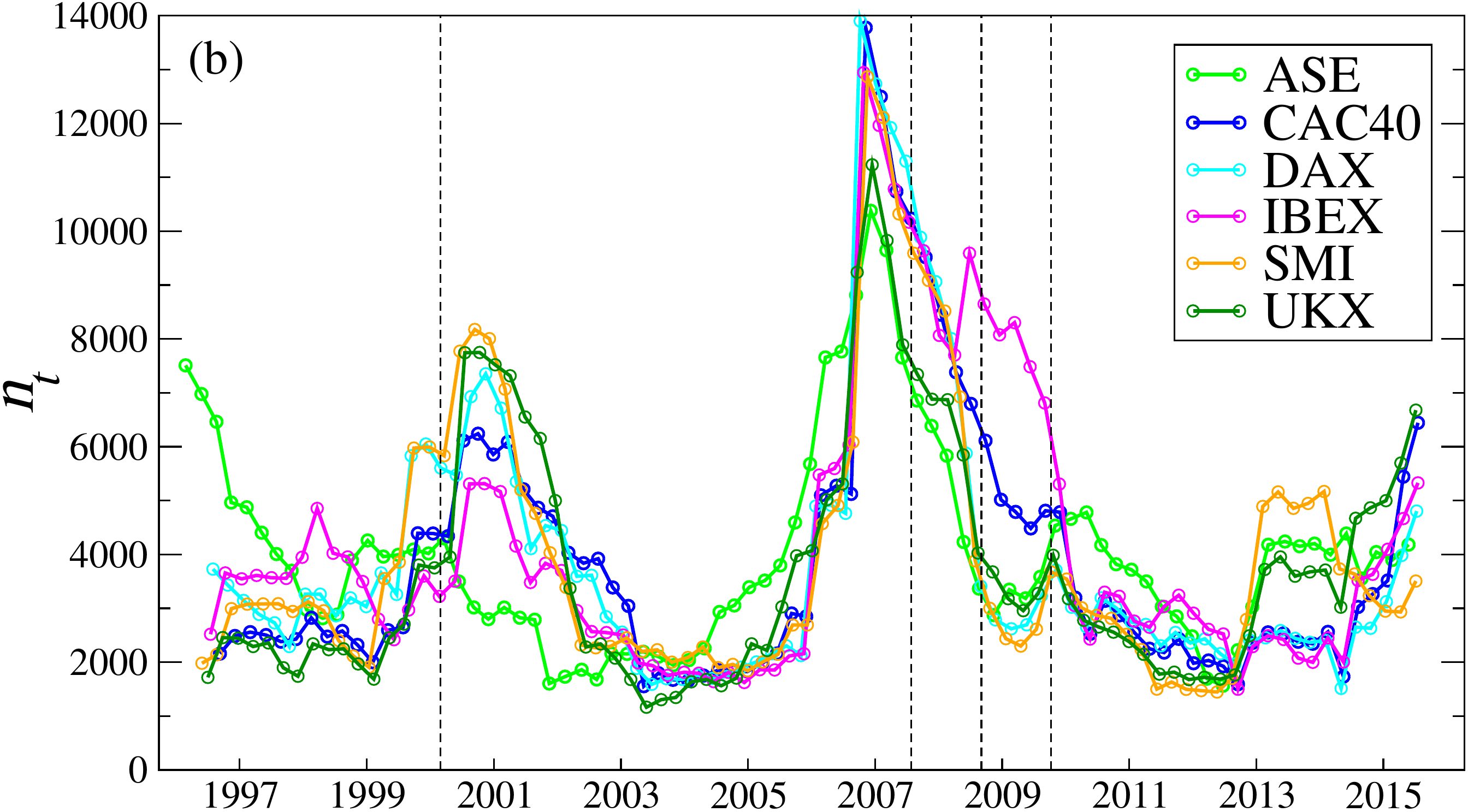}
\caption{Number of validated links for S\&P500 and MERVAL - American countries, panel (a) - and for ASE, CAC40, DAX, IBEX, SMI, UKX - European countries, panel (b). 
Vertical dashed lines indicate dates of relevant events. 
In chronological order: 13 March 2000, the day after the dot-com bubble peaked and prices started to fall; 9 August 2007, when BNP Paribas froze three of their funds on sub-prime mortgage markets; 
15 September 2008, when Lehman Brothers filed for bankruptcy; 17 October 2009, when Greece unveiled the amount of holes in their finances, initiating the Euro sovereign debt crisis.}\label{fig3}
\end{figure*}

We now use the proposed validation methodology to look for time series patterns related to unstable market periods. 
In order to detect temporal changes that cannot be described by the null model, we use a sliding window technique. 
Let $W$ denote the window length, and $L$ the shift amount. Then, for each $\lambda=0,1,\dots,\lfloor (T-W)/L \rfloor$, 
we take the elements $\{\sigma_t\}_{t=\lambda L}^{W+\lambda L}$ of the empirical time series, create the corresponding empirical and null model VG, 
apply the validation procedure within these subgraphs, and compute the number of validated connections---which are denoted as $n_{W+\lambda L}$. 
The result of this procedure is a time series
\begin{equation}
n_W,\,n_{W+L},\,n_{W+2L},\,\dots,
\end{equation} 
where the time stamp of each element corresponds to the last date of the corresponding time window: we use only past information to compute the number of validated connections.
In the following we set $W=500$ (two financial years), $L=60$ (three financial months) and $\rho= 0.1$. 
Results and discussion for different choices of parameters are reported in the appendix, Figure \ref{fig5}.

Figure \ref{fig3} shows the number of validated connections at different dates and for the various global market indices considered. 
Focusing first on S\&P500, we observe five local maxima, each starting at the onset of a crisis period. 
In chronological order we find the Argentine great depression (1998-2002), the dot-com bubble burst (2000-2002), the sub-prime mortgage crisis (2007-2008), 
the Euro sovereign debt crisis (since 2009), and the Russian financial crisis (2014-2017).
Remarkably, we see that in all cases $n$ starts to increase {\em before} the crisis occurs. 
This is particularly evident in the case of the sub-prime crisis, as $n$ started a rapid ascent since the summer of 2006 (\ie, one year before the crisis). 
Thus, the number of validated connections turns out to provide early warnings of financial turmoil periods.

Moving to the other indices, all of them feature patterns similar to that of S\&P500, with due peculiarities. 
In the case of MERVAL (Argentine), for instance, we find a more pronounced peak at the beginning of the Argentine depression, and an even more marked one 
just afterwards--- which is supposedly more related to the 2002 crisis of the Argentine peso than to the dot-com bubble burst.
ASE (Greece) features three distinctive traits: i) a peak around 1996, probably related to the Greek-Turkish crisis occurring in those years; 
ii) an increasing trend towards the sub-prime crisis starting as early as 2004 (\ie, before that of all other indices considered), 
when the country faced important organizational expenses for the Olympic Games; iii) a peak before the Greek debt crisis.
IBEX (Spain) is the only European index affected by the Argentine depression, and features an additional peak in 2008 as Spain was the European country 
most affected by the housing bubble, causing an early economic recession. CAC40 (France), DAX (Germany), SMI (Switzerland) and UKX (United Kingdom) do not show particular trends, 
with the exception of SMI featuring a marked peak starting in 2013---in the midst of the period when Swiss Franc was pegged to the Euro.

\subsection{Validated visibility}

As far as now, we have obtained a qualitative indicator with the ability to mark and forecast financial instabilities. 
Yet, our validation framework allows to get more quantitative insights. Using the sliding window technique described above, 
we define the {\em validated visibility} for time window $t$ as: 
\begin{equation}
V_t=(n_t/\avg{d}_t)\Big/(\bar{n}_t/\avg{\bar{d}}_t),\label{eq:vv}
\end{equation}
namely the ratio between the number of validated links over the average connectivity in the empirical VG and the number of validated links over the average connectivity in the empirical IVG.
The rationale of this index is the following. For a market trend which is properly described by the null model, 
the likelihood to validate a connection is the same for VG and IVG. As such, $V\simeq 1$. During a financial bubble, instead, 
the super-exponential growth of market prices \cite{sornette2003critical} leads to a drastic increase in the number of VG connections \cite{yan2015forecasting}, 
which are mostly validated so that $n/\avg{d}$ increases. At the same time, IVG connectivity decreases and validation becomes less likely (recall that missing connections cannot be validated), 
so that $\bar{n}/\avg{\bar{d}}$ decreases. The opposite pattern occurs during negative bubbles \cite{yan2012diagnosis,yan2015forecasting}. 
Overall, $V>1$ signals a speculative bubble, often but not always followed by a market crash, whereas, $V<1$ signals a negative bubble, in general followed by large rebounds or rallies.
$V=1$ is thus the critical value separating these two regimes.

\begin{figure*}[t]
\centering
\includegraphics[width=0.48\textwidth]{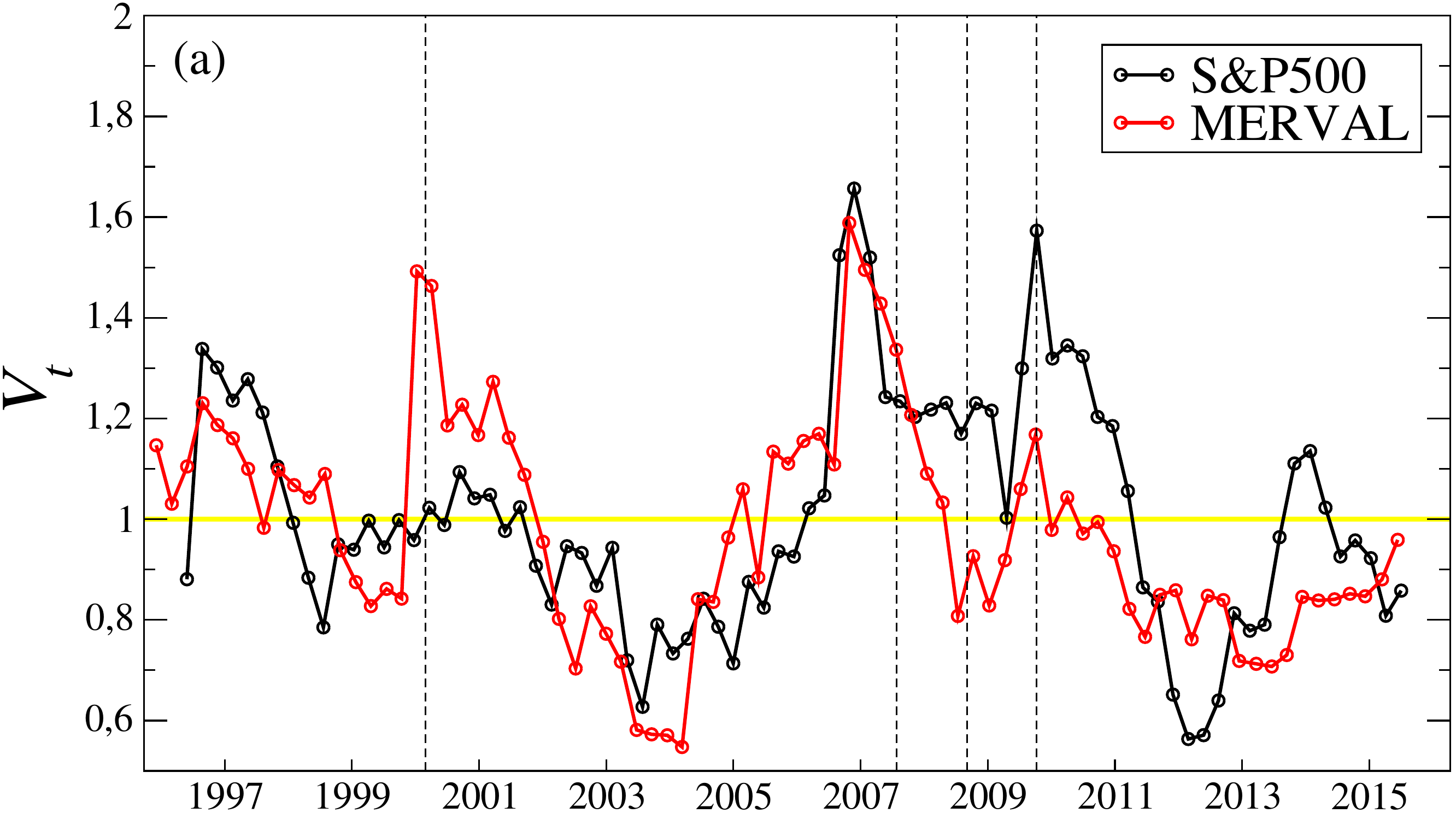}
\includegraphics[width=0.48\textwidth]{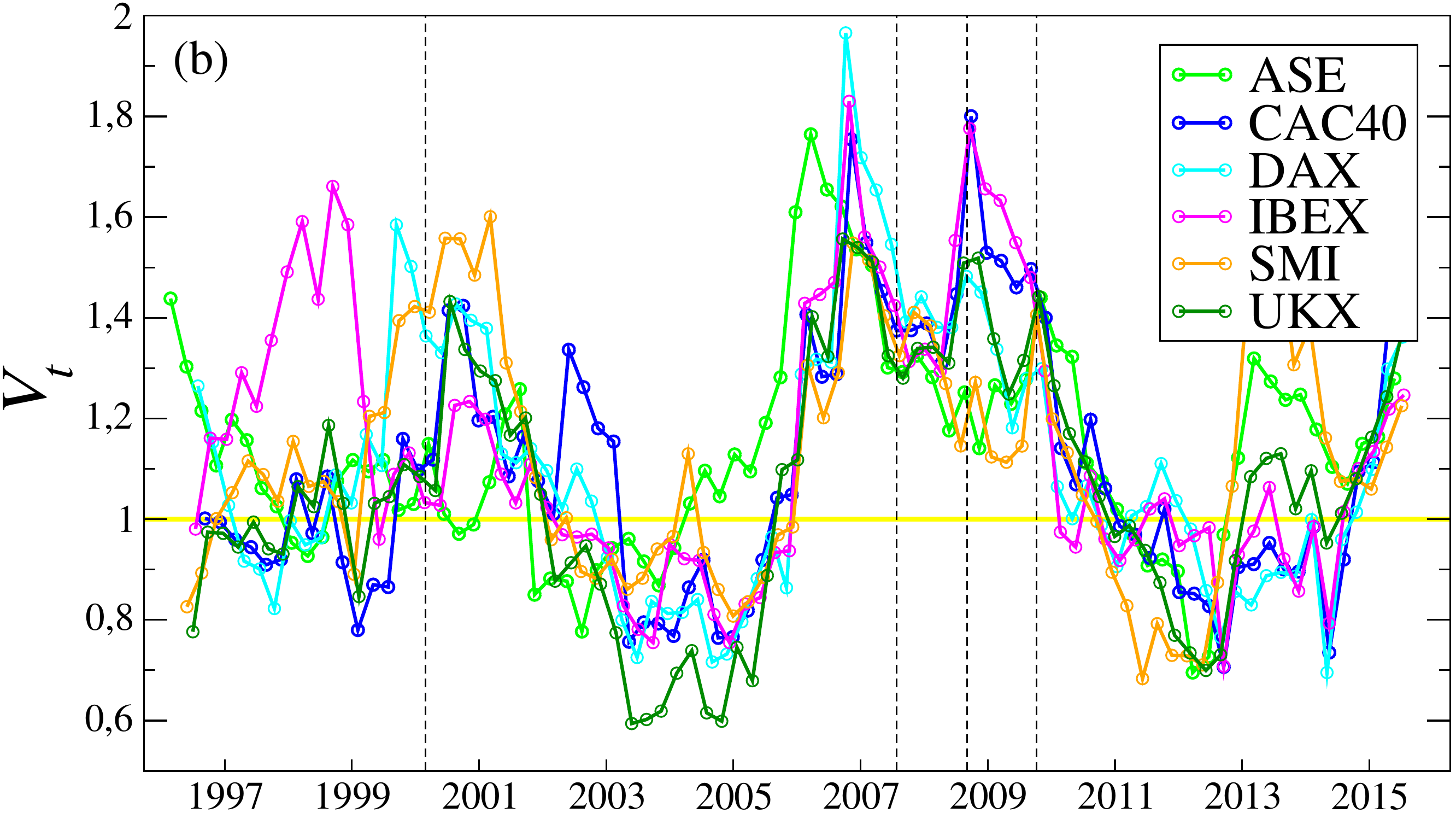}
\caption{Validated visibility for S\&P500 and MERVAL - American countries, panel (a) - and for ASE, CAC40, DAX, IBEX, SMI, UKX - European countries, panel (b). 
The horizontal solid line marks the critical value of 1 above which financial instability occurs. Vertical dashed lines indicate the same particular dates as Figure \ref{fig3}.}\label{fig4}
\end{figure*}

Figure \ref{fig4} shows how validated visibility evolves over time. In the case of S\&P500, $V$ is systematically higher than one in correspondence of all crisis periods, 
and in general the signal is magnified with respect to what we observed for $n$. In particular, concerning the peak of the global financial crisis, 
$V$ goes beyond the critical value as far ahead as early 2006. Similar patterns are observed for the other market indices. 
Overall, $V$ comes forth as a robust quantitative indicator providing early warnings of financial instability periods. 

\section{Conclusions}

In this work we developed a statistical validation technique for financial time series mapped into networks 
via the visibility graph algorithm \cite{lacasa2008from}. We use ARCH-type models, commonly employed to model and forecast assets price series, to build the null hypothesis, 
with the underlying idea that such models cannot capture super-exponential patterns observed during bubbles and before crashes. 
By applying our methodology to global market indices series, we showed how to extract a quantitative signal providing early warnings of financial crisis, but also of local instabilities. 
Furthermore, the extent of the crisis is captured by the intensity of the signal. As such, our findings can have potential application 
for practitioners of portfolio optimization and hedging, and more generally for regulators to measure market efficiency and foresee systemic events.

We finally remark that our validation framework is rather flexible. For instance, we can use a dynamical null model for which GJR-GARCH parameters are estimated within each sliding time window. 
Additionally, by choosing an appropriate null model, our method can be applied to other financial instruments, such as derivatives, CDS, and so forth. 
The validity of the strategy behind the proposed validation procedure for time-series goes however well beyond the financial domain. 
Indeed, our validation recipe can be applied for studying features of time-series of any nature, provided a suitable null model can be defined for them. 
And when such a null model is not specified, it is possible to resort to standard network randomization schemes (configuration models or exponential random graphs) 
\cite{newman2001random,park2004statistical,squartini2011analytical} which preserve local topological properties of the VG.


This work was supported by the EU projects DOLFINS (grant 640772) and CoeGSS (grant 676547), and the Italian PNR project CRISIS-Lab. 
The funders had no role in study design, data collection and analysis, decision to publish, or preparation of the manuscript.

\newpage

\appendix

\begin{table*}[t!]
\caption{GJR-GARCH parameters estimation for the considered equity indices.}\label{tab:other}
\begin{tabular}{c c c c c}
\hline
\textbf{parameters} & \textbf{estimate} & \textbf{std. error} & \textbf{$t$-statistic} & \textbf{$p$-value}\\ 
\hline\hline
\multicolumn{5}{c}{MERVAL}\\ \hline
$\alpha_{0}$ & 0.12  & 0.023   & 5.4   & $<10^{-7}$\\
$\alpha_{1}$ & 0.041  &  0.009  &  4.5  & $<10^{-5}$ \\
$\beta_{1} $ & 0.86 &  0.01  & 73.5 & $0^*$\\
$\gamma_{1}$ &  0.13  &  0.01   & 8.1  & $<10^{-13}$ \\
dof        &  5.7  &  0.4    &  13.6 &  $0^*$ \\
\hline\hline
\multicolumn{5}{c}{ASE}\\ \hline
$\alpha_{0}$ & 0.029  & 0.007   & 4.0   & 0.0001\\
$\alpha_{1}$ & 0.063  & 0.008   & 7.1    & $<10^{-12}$\\
$\beta_{1} $ & 0.910 &  0.007   & 124.0  & $0^*$\\
$\gamma_{1}$ & 0.04  &  0.01    & 3.6    & $<10^{-4}$\\
dof        &  6.7  &  0.6    &  12.1    &  $0^*$ \\
\hline\hline
\multicolumn{5}{c}{CAC40}\\ \hline
$\alpha_{0}$ & 0.014  & 0.004   & 3.4   & 0.0007\\
$\alpha_{1}$ & 0.018  & 0.007   & 2.7    & 0.0062\\
$\beta_{1} $ & 0.926 &  0.007   & 132.5  & $0^*$\\
$\gamma_{1}$ & 0.09  &  0.01    & 8.1    & $<10^{-15}$\\
dof        &  9.8  &  1.0    &  8.9    &  $0^*$ \\
\hline\hline
\multicolumn{5}{c}{DAX}\\ \hline
$\alpha_{0}$ & 0.010  & 0.004   & 3.9   & 0.0040\\
$\alpha_{1}$ & 0.025  & 0.007   & 3.5    & 0.0005\\
$\beta_{1} $ & 0.923 &  0.007   & 133.3  & $0^*$\\
$\gamma_{1}$ & 0.10  &  0.01    & 8.0    & $<10^{-14}$\\
dof        &  9.2  &  1.0    &  8.8    &  $0^*$ \\
\hline\hline
\multicolumn{5}{c}{IBEX}\\ \hline
$\alpha_{0}$ & 0.015  & 0.004   & 3.44   &  0.0006 \\
$\alpha_{1}$ & 0.024  &  0.007  &  3.5  & 0.0005\\
$\beta_{1} $ & 0.923 &  0.007  & 127.4 & $0^*$\\
$\gamma_{1}$ &  0.09  &  0.01   & 8.1  & $<10^{-15}$\\
dof        &  9.5  &  0.9    &  9.7 &  $0^*$ \\
\hline\hline
\multicolumn{5}{c}{SMI}\\ \hline
$\alpha_{0}$ & 0.018  & 0.004   & 4.5   &  $<10^{-5}$\\
$\alpha_{1}$ & 0.040  &  0.008  &  4.7   &  $<10^{-5}$ \\
$\beta_{1} $ & 0.905  &  0.009  & 103.6 & $0^*$\\
$\gamma_{1}$ &  0.083  &  0.01   & 6.7  & $<10^{-10}$ \\
dof        &  11.5  &  1.5    &  7.3 &  $<10^{-12}$ \\
\hline\hline
\multicolumn{5}{c}{UKX}\\ \hline
$\alpha_{0}$ & 0.010  & 0.003   & 3.5    & 0.0004 \\
$\alpha_{1}$ & 0.022  &  0.007  &  2.9   &  0.0037 \\
$\beta_{1} $ & 0.916  &  0.007  & 125.1 & $0^*$\\
$\gamma_{1}$ &  0.11  &  0.01  & 8.5  & $0^*$ \\
dof        &  11.2  &  1.6   &  6.9 &  $<10^{-11}$ \\
\hline
\end{tabular}
\end{table*}

\begin{figure*}[t!]
\centering
\includegraphics[width=\textwidth]{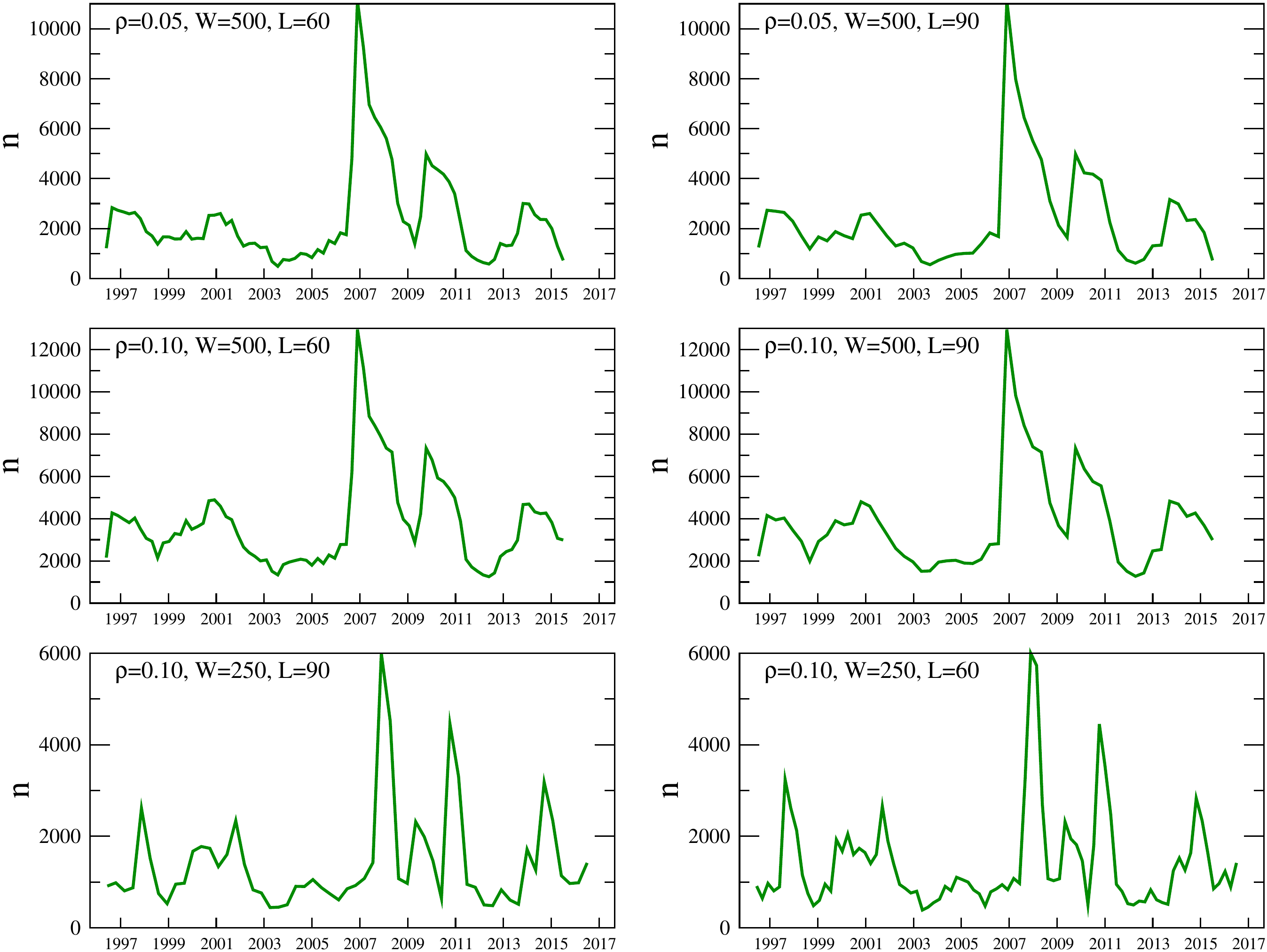}
\caption{Number of VG validated links for S\&P500 using different values of $\rho$ (validation threshold), $W$ (window length) and $L$ (shift amount).
We see that results are robust with respect to the choice of the threshold: as $\rho$ changes, only the intensity of the signal changes but not its shape. 
Yet, $\rho$ must be not too high to harm the specificity of the test (in the limiting case $\rho\to 1$ all connections are validated), 
nor too low to get low sensitivity (for $\rho\to 0$ no connection is validated and, according to the considerations of Figure\ref{fig2}(b), 
the minimum value allowed for the threshold is set by the window length $W$). 
Concerning $W$ and $L$, generally increasing their values reduces the noise of the signal but also its temporal resolution, and vice-versa. 
The most appropriate balance between $W$ and $L$ thus appears to be application-specific.}\label{fig5}
\end{figure*}


\begin{thebibliography}{41}%
\makeatletter
\providecommand \@ifxundefined [1]{%
 \@ifx{#1\undefined}
}%
\providecommand \@ifnum [1]{%
 \ifnum #1\expandafter \@firstoftwo
 \else \expandafter \@secondoftwo
 \fi
}%
\providecommand \@ifx [1]{%
 \ifx #1\expandafter \@firstoftwo
 \else \expandafter \@secondoftwo
 \fi
}%
\providecommand \natexlab [1]{#1}%
\providecommand \enquote  [1]{``#1''}%
\providecommand \bibnamefont  [1]{#1}%
\providecommand \bibfnamefont [1]{#1}%
\providecommand \citenamefont [1]{#1}%
\providecommand \href@noop [0]{\@secondoftwo}%
\providecommand \href [0]{\begingroup \@sanitize@url \@href}%
\providecommand \@href[1]{\@@startlink{#1}\@@href}%
\providecommand \@@href[1]{\endgroup#1\@@endlink}%
\providecommand \@sanitize@url [0]{\catcode `\\12\catcode `\$12\catcode
  `\&12\catcode `\#12\catcode `\^12\catcode `\_12\catcode `\%12\relax}%
\providecommand \@@startlink[1]{}%
\providecommand \@@endlink[0]{}%
\providecommand \url  [0]{\begingroup\@sanitize@url \@url }%
\providecommand \@url [1]{\endgroup\@href {#1}{\urlprefix }}%
\providecommand \urlprefix  [0]{URL }%
\providecommand \Eprint [0]{\href }%
\providecommand \doibase [0]{http://dx.doi.org/}%
\providecommand \selectlanguage [0]{\@gobble}%
\providecommand \bibinfo  [0]{\@secondoftwo}%
\providecommand \bibfield  [0]{\@secondoftwo}%
\providecommand \translation [1]{[#1]}%
\providecommand \BibitemOpen [0]{}%
\providecommand \bibitemStop [0]{}%
\providecommand \bibitemNoStop [0]{.\EOS\space}%
\providecommand \EOS [0]{\spacefactor3000\relax}%
\providecommand \BibitemShut  [1]{\csname bibitem#1\endcsname}%
\let\auto@bib@innerbib\@empty
\bibitem [{\citenamefont {Caldarelli}(2007)}]{caldarelli2007scale-free}%
  \BibitemOpen
  \bibfield  {author} {\bibinfo {author} {\bibfnamefont {G.}~\bibnamefont
  {Caldarelli}},\ }\href {\doibase 10.1093/acprof:oso/9780199211517.001.0001}
  {\emph {\bibinfo {title} {Scale-Free Networks: Complex Webs in Nature and
  Technology}}}\ (\bibinfo  {publisher} {Oxford University Press},\ \bibinfo
  {year} {2007})\BibitemShut {NoStop}%
\bibitem [{\citenamefont {Cimini}\ \emph {et~al.}(2015)\citenamefont {Cimini},
  \citenamefont {Squartini}, \citenamefont {Garlaschelli},\ and\ \citenamefont
  {Gabrielli}}]{cimini2015systemic}%
  \BibitemOpen
  \bibfield  {author} {\bibinfo {author} {\bibfnamefont {G.}~\bibnamefont
  {Cimini}}, \bibinfo {author} {\bibfnamefont {T.}~\bibnamefont {Squartini}},
  \bibinfo {author} {\bibfnamefont {D.}~\bibnamefont {Garlaschelli}}, \ and\
  \bibinfo {author} {\bibfnamefont {A.}~\bibnamefont {Gabrielli}},\ }\href
  {\doibase 10.1038/srep15758} {\bibfield  {journal} {\bibinfo  {journal}
  {Scientific Reports}\ }\textbf {\bibinfo {volume} {5}},\ \bibinfo {pages}
  {15758} (\bibinfo {year} {2015})}\BibitemShut {NoStop}%
\bibitem [{\citenamefont {Straka}\ \emph {et~al.}(2017)\citenamefont {Straka},
  \citenamefont {Caldarelli},\ and\ \citenamefont {Saracco}}]{straka2017grand}%
  \BibitemOpen
  \bibfield  {author} {\bibinfo {author} {\bibfnamefont {M.~J.}\ \bibnamefont
  {Straka}}, \bibinfo {author} {\bibfnamefont {G.}~\bibnamefont {Caldarelli}},
  \ and\ \bibinfo {author} {\bibfnamefont {F.}~\bibnamefont {Saracco}},\ }\href
  {\doibase 10.1103/PhysRevE.96.022306} {\bibfield  {journal} {\bibinfo
  {journal} {Physical Review E}\ }\textbf {\bibinfo {volume} {96}},\ \bibinfo
  {pages} {022306} (\bibinfo {year} {2017})}\BibitemShut {NoStop}%
\bibitem [{\citenamefont {Mishkin}(1999)}]{mishkin1999global}%
  \BibitemOpen
  \bibfield  {author} {\bibinfo {author} {\bibfnamefont {F.~S.}\ \bibnamefont
  {Mishkin}},\ }\href {\doibase 10.1257/jep.13.4.3} {\bibfield  {journal}
  {\bibinfo  {journal} {The Journal of Economic Perspectives}\ }\textbf
  {\bibinfo {volume} {13}},\ \bibinfo {pages} {3} (\bibinfo {year}
  {1999})}\BibitemShut {NoStop}%
\bibitem [{\citenamefont {Tsay}(2005)}]{tsay2005analysis}%
  \BibitemOpen
  \bibfield  {author} {\bibinfo {author} {\bibfnamefont {R.~S.}\ \bibnamefont
  {Tsay}},\ }\href@noop {} {\emph {\bibinfo {title} {Analysis of Financial Time
  Series}}},\ Vol.\ \bibinfo {volume} {543}\ (\bibinfo  {publisher} {John Wiley
  \& Sons},\ \bibinfo {year} {2005})\BibitemShut {NoStop}%
\bibitem [{\citenamefont {Chakraborti}\ \emph {et~al.}(2011)\citenamefont
  {Chakraborti}, \citenamefont {M}, \citenamefont {M},\ and\ \citenamefont
  {Abergel}}]{chakraborti2011econophysics}%
  \BibitemOpen
  \bibfield  {author} {\bibinfo {author} {\bibfnamefont {A.}~\bibnamefont
  {Chakraborti}}, \bibinfo {author} {\bibfnamefont {T.~I.}\ \bibnamefont {M}},
  \bibinfo {author} {\bibfnamefont {P.}~\bibnamefont {M}}, \ and\ \bibinfo
  {author} {\bibfnamefont {F.}~\bibnamefont {Abergel}},\ }\href {\doibase
  10.1080/14697688.2010.539248} {\bibfield  {journal} {\bibinfo  {journal}
  {Quantitative Finance}\ }\textbf {\bibinfo {volume} {11}},\ \bibinfo {pages}
  {991} (\bibinfo {year} {2011})}\BibitemShut {NoStop}%
\bibitem [{\citenamefont {Malkiel}\ and\ \citenamefont
  {Fama}(1970)}]{malkiel1970efficient}%
  \BibitemOpen
  \bibfield  {author} {\bibinfo {author} {\bibfnamefont {B.~G.}\ \bibnamefont
  {Malkiel}}\ and\ \bibinfo {author} {\bibfnamefont {E.~F.}\ \bibnamefont
  {Fama}},\ }\href {\doibase 10.1111/j.1540-6261.1970.tb00518.x} {\bibfield
  {journal} {\bibinfo  {journal} {The Journal of Finance}\ }\textbf {\bibinfo
  {volume} {25}},\ \bibinfo {pages} {383} (\bibinfo {year} {1970})}\BibitemShut
  {NoStop}%
\bibitem [{\citenamefont {Cont}(2001)}]{cont2001empirical}%
  \BibitemOpen
  \bibfield  {author} {\bibinfo {author} {\bibfnamefont {R.}~\bibnamefont
  {Cont}},\ }\href {\doibase 10.1080/713665670} {\bibfield  {journal} {\bibinfo
   {journal} {Quantitative Finance}\ }\textbf {\bibinfo {volume} {1}},\
  \bibinfo {pages} {223} (\bibinfo {year} {2001})}\BibitemShut {NoStop}%
\bibitem [{\citenamefont {Engle}(1982)}]{engle1982autoregressive}%
  \BibitemOpen
  \bibfield  {author} {\bibinfo {author} {\bibfnamefont {R.~F.}\ \bibnamefont
  {Engle}},\ }\href {\doibase 10.2307/1912773} {\bibfield  {journal} {\bibinfo
  {journal} {Econometrica: Journal of the Econometric Society}\ }\textbf
  {\bibinfo {volume} {50}},\ \bibinfo {pages} {987} (\bibinfo {year}
  {1982})}\BibitemShut {NoStop}%
\bibitem [{\citenamefont {Bollerslev}(1986)}]{bollerslev1986generalized}%
  \BibitemOpen
  \bibfield  {author} {\bibinfo {author} {\bibfnamefont {T.}~\bibnamefont
  {Bollerslev}},\ }\href {\doibase 10.1016/0304-4076(86)90063-1} {\bibfield
  {journal} {\bibinfo  {journal} {Journal of Econometrics}\ }\textbf {\bibinfo
  {volume} {31}},\ \bibinfo {pages} {307} (\bibinfo {year} {1986})}\BibitemShut
  {NoStop}%
\bibitem [{\citenamefont {Glosten}\ \emph {et~al.}(1993)\citenamefont
  {Glosten}, \citenamefont {Jagannathan},\ and\ \citenamefont
  {Runkle}}]{glosten1993relation}%
  \BibitemOpen
  \bibfield  {author} {\bibinfo {author} {\bibfnamefont {L.~R.}\ \bibnamefont
  {Glosten}}, \bibinfo {author} {\bibfnamefont {R.}~\bibnamefont
  {Jagannathan}}, \ and\ \bibinfo {author} {\bibfnamefont {D.~E.}\ \bibnamefont
  {Runkle}},\ }\href {\doibase 10.1111/j.1540-6261.1993.tb05128.x} {\bibfield
  {journal} {\bibinfo  {journal} {The Journal of Finance}\ }\textbf {\bibinfo
  {volume} {48}},\ \bibinfo {pages} {1779} (\bibinfo {year}
  {1993})}\BibitemShut {NoStop}%
\bibitem [{\citenamefont {Gao}\ \emph {et~al.}(2016)\citenamefont {Gao},
  \citenamefont {Small},\ and\ \citenamefont {Kurths}}]{gao2016complex}%
  \BibitemOpen
  \bibfield  {author} {\bibinfo {author} {\bibfnamefont {Z.-K.}\ \bibnamefont
  {Gao}}, \bibinfo {author} {\bibfnamefont {M.}~\bibnamefont {Small}}, \ and\
  \bibinfo {author} {\bibfnamefont {J.}~\bibnamefont {Kurths}},\ }\href
  {\doibase 10.1209/0295-5075/116/50001} {\bibfield  {journal} {\bibinfo
  {journal} {Europhysics Letters}\ }\textbf {\bibinfo {volume} {116}},\
  \bibinfo {pages} {50001} (\bibinfo {year} {2016})}\BibitemShut {NoStop}%
\bibitem [{\citenamefont {Doye}(2002)}]{doye2002network}%
  \BibitemOpen
  \bibfield  {author} {\bibinfo {author} {\bibfnamefont {J.~P.}\ \bibnamefont
  {Doye}},\ }\href {\doibase 10.1103/PhysRevLett.88.238701} {\bibfield
  {journal} {\bibinfo  {journal} {Physical Review Letters}\ }\textbf {\bibinfo
  {volume} {88}},\ \bibinfo {pages} {238701} (\bibinfo {year}
  {2002})}\BibitemShut {NoStop}%
\bibitem [{\citenamefont {Gfeller}\ \emph {et~al.}(2007)\citenamefont
  {Gfeller}, \citenamefont {De~Los~Rios}, \citenamefont {Caflisch},\ and\
  \citenamefont {Rao}}]{gfeller2007complex}%
  \BibitemOpen
  \bibfield  {author} {\bibinfo {author} {\bibfnamefont {D.}~\bibnamefont
  {Gfeller}}, \bibinfo {author} {\bibfnamefont {P.}~\bibnamefont
  {De~Los~Rios}}, \bibinfo {author} {\bibfnamefont {A.}~\bibnamefont
  {Caflisch}}, \ and\ \bibinfo {author} {\bibfnamefont {F.}~\bibnamefont
  {Rao}},\ }\href {\doibase 10.1073/pnas.0608099104} {\bibfield  {journal}
  {\bibinfo  {journal} {PNAS}\ }\textbf {\bibinfo {volume} {104}},\ \bibinfo
  {pages} {1817} (\bibinfo {year} {2007})}\BibitemShut {NoStop}%
\bibitem [{\citenamefont {Lacasa}\ \emph {et~al.}(2008)\citenamefont {Lacasa},
  \citenamefont {Luque}, \citenamefont {Ballesteros}, \citenamefont {Luque},\
  and\ \citenamefont {Nuno}}]{lacasa2008from}%
  \BibitemOpen
  \bibfield  {author} {\bibinfo {author} {\bibfnamefont {L.}~\bibnamefont
  {Lacasa}}, \bibinfo {author} {\bibfnamefont {B.}~\bibnamefont {Luque}},
  \bibinfo {author} {\bibfnamefont {F.}~\bibnamefont {Ballesteros}}, \bibinfo
  {author} {\bibfnamefont {J.}~\bibnamefont {Luque}}, \ and\ \bibinfo {author}
  {\bibfnamefont {J.~C.}\ \bibnamefont {Nuno}},\ }\href {\doibase
  10.1073/pnas.0709247105} {\bibfield  {journal} {\bibinfo  {journal} {PNAS}\
  }\textbf {\bibinfo {volume} {105}},\ \bibinfo {pages} {4972} (\bibinfo {year}
  {2008})}\BibitemShut {NoStop}%
\bibitem [{\citenamefont {Nu{\~n}ez}\ \emph {et~al.}(2012)\citenamefont
  {Nu{\~n}ez}, \citenamefont {Lacasa}, \citenamefont {Gomez},\ and\
  \citenamefont {Luque}}]{nunez2012visibility}%
  \BibitemOpen
  \bibfield  {author} {\bibinfo {author} {\bibfnamefont {A.~M.}\ \bibnamefont
  {Nu{\~n}ez}}, \bibinfo {author} {\bibfnamefont {L.}~\bibnamefont {Lacasa}},
  \bibinfo {author} {\bibfnamefont {J.~P.}\ \bibnamefont {Gomez}}, \ and\
  \bibinfo {author} {\bibfnamefont {B.}~\bibnamefont {Luque}},\ }\enquote
  {\bibinfo {title} {Visibility algorithms: A short review},}\ in\ \href
  {\doibase 10.5772/34810} {\emph {\bibinfo {booktitle} {New Frontiers in Graph
  Theory}}}\ (\bibinfo  {publisher} {InTech},\ \bibinfo {year}
  {2012})\BibitemShut {NoStop}%
\bibitem [{\citenamefont {Luque}\ \emph {et~al.}(2009)\citenamefont {Luque},
  \citenamefont {Lacasa}, \citenamefont {Ballesteros},\ and\ \citenamefont
  {Luque}}]{luque2009horizontal}%
  \BibitemOpen
  \bibfield  {author} {\bibinfo {author} {\bibfnamefont {B.}~\bibnamefont
  {Luque}}, \bibinfo {author} {\bibfnamefont {L.}~\bibnamefont {Lacasa}},
  \bibinfo {author} {\bibfnamefont {F.}~\bibnamefont {Ballesteros}}, \ and\
  \bibinfo {author} {\bibfnamefont {J.}~\bibnamefont {Luque}},\ }\href
  {\doibase 10.1103/PhysRevE.80.046103} {\bibfield  {journal} {\bibinfo
  {journal} {Physical Review E}\ }\textbf {\bibinfo {volume} {80}},\ \bibinfo
  {pages} {046103} (\bibinfo {year} {2009})}\BibitemShut {NoStop}%
\bibitem [{\citenamefont {Lacasa}\ and\ \citenamefont
  {Toral}(2010)}]{lacasa2010description}%
  \BibitemOpen
  \bibfield  {author} {\bibinfo {author} {\bibfnamefont {L.}~\bibnamefont
  {Lacasa}}\ and\ \bibinfo {author} {\bibfnamefont {R.}~\bibnamefont {Toral}},\
  }\href {\doibase 10.1103/PhysRevE.82.036120} {\bibfield  {journal} {\bibinfo
  {journal} {Physical Review E}\ }\textbf {\bibinfo {volume} {82}},\ \bibinfo
  {pages} {036120} (\bibinfo {year} {2010})}\BibitemShut {NoStop}%
\bibitem [{\citenamefont {Lacasa}\ \emph {et~al.}(2015)\citenamefont {Lacasa},
  \citenamefont {Nicosia},\ and\ \citenamefont {Latora}}]{lacasa2015network}%
  \BibitemOpen
  \bibfield  {author} {\bibinfo {author} {\bibfnamefont {L.}~\bibnamefont
  {Lacasa}}, \bibinfo {author} {\bibfnamefont {V.}~\bibnamefont {Nicosia}}, \
  and\ \bibinfo {author} {\bibfnamefont {V.}~\bibnamefont {Latora}},\ }\href
  {\doibase 10.1038/srep15508} {\bibfield  {journal} {\bibinfo  {journal}
  {Scientific Reports}\ }\textbf {\bibinfo {volume} {5}},\ \bibinfo {pages}
  {15508} (\bibinfo {year} {2015})}\BibitemShut {NoStop}%
\bibitem [{\citenamefont {Stephen}\ \emph {et~al.}(2015)\citenamefont
  {Stephen}, \citenamefont {Gu},\ and\ \citenamefont
  {Yang}}]{stephen2015visibility}%
  \BibitemOpen
  \bibfield  {author} {\bibinfo {author} {\bibfnamefont {M.}~\bibnamefont
  {Stephen}}, \bibinfo {author} {\bibfnamefont {C.}~\bibnamefont {Gu}}, \ and\
  \bibinfo {author} {\bibfnamefont {H.}~\bibnamefont {Yang}},\ }\href {\doibase
  10.1371/journal.pone.0143015} {\bibfield  {journal} {\bibinfo  {journal}
  {PLoS ONE}\ }\textbf {\bibinfo {volume} {10}},\ \bibinfo {pages} {e0143015}
  (\bibinfo {year} {2015})}\BibitemShut {NoStop}%
\bibitem [{\citenamefont {Yang}\ \emph {et~al.}(2009)\citenamefont {Yang},
  \citenamefont {Wang}, \citenamefont {Yang},\ and\ \citenamefont
  {Mang}}]{yang2009visibility}%
  \BibitemOpen
  \bibfield  {author} {\bibinfo {author} {\bibfnamefont {Y.}~\bibnamefont
  {Yang}}, \bibinfo {author} {\bibfnamefont {J.}~\bibnamefont {Wang}}, \bibinfo
  {author} {\bibfnamefont {H.}~\bibnamefont {Yang}}, \ and\ \bibinfo {author}
  {\bibfnamefont {J.}~\bibnamefont {Mang}},\ }\href {\doibase
  10.1016/j.physa.2009.07.016} {\bibfield  {journal} {\bibinfo  {journal}
  {Physica A: Statistical Mechanics and its Applications}\ }\textbf {\bibinfo
  {volume} {388}},\ \bibinfo {pages} {4431} (\bibinfo {year}
  {2009})}\BibitemShut {NoStop}%
\bibitem [{\citenamefont {Rasheed}\ and\ \citenamefont
  {Qian}(2004)}]{rasheed2004hurst}%
  \BibitemOpen
  \bibfield  {author} {\bibinfo {author} {\bibfnamefont {B.~Q.~K.}\
  \bibnamefont {Rasheed}}\ and\ \bibinfo {author} {\bibfnamefont
  {B.}~\bibnamefont {Qian}},\ }in\ \href@noop {} {\emph {\bibinfo {booktitle}
  {2nd IASTED intl. conference on Financial Engineering and Applications}}}\
  (\bibinfo {year} {2004})\ pp.\ \bibinfo {pages} {203--209}\BibitemShut
  {NoStop}%
\bibitem [{\citenamefont {Yan}\ and\ \citenamefont {van
  Serooskerken}(2015)}]{yan2015forecasting}%
  \BibitemOpen
  \bibfield  {author} {\bibinfo {author} {\bibfnamefont {W.}~\bibnamefont
  {Yan}}\ and\ \bibinfo {author} {\bibfnamefont {E.~v.~T.}\ \bibnamefont {van
  Serooskerken}},\ }\href {\doibase 10.1371/journal.pone.0128908} {\bibfield
  {journal} {\bibinfo  {journal} {PLoS ONE}\ }\textbf {\bibinfo {volume}
  {10}},\ \bibinfo {pages} {e0128908} (\bibinfo {year} {2015})}\BibitemShut
  {NoStop}%
\bibitem [{\citenamefont {Zhang}\ \emph {et~al.}(2017)\citenamefont {Zhang},
  \citenamefont {Zou}, \citenamefont {Zhou}, \citenamefont {Gao},\ and\
  \citenamefont {Guan}}]{zhang2017visibility}%
  \BibitemOpen
  \bibfield  {author} {\bibinfo {author} {\bibfnamefont {R.}~\bibnamefont
  {Zhang}}, \bibinfo {author} {\bibfnamefont {Y.}~\bibnamefont {Zou}}, \bibinfo
  {author} {\bibfnamefont {J.}~\bibnamefont {Zhou}}, \bibinfo {author}
  {\bibfnamefont {Z.-K.}\ \bibnamefont {Gao}}, \ and\ \bibinfo {author}
  {\bibfnamefont {S.}~\bibnamefont {Guan}},\ }\href {\doibase
  10.1016/j.cnsns.2016.04.031} {\bibfield  {journal} {\bibinfo  {journal}
  {Communications in Nonlinear Science and Numerical Simulation}\ }\textbf
  {\bibinfo {volume} {42}},\ \bibinfo {pages} {396} (\bibinfo {year}
  {2017})}\BibitemShut {NoStop}%
\bibitem [{\citenamefont {Gon\c{c}alves}\ \emph {et~al.}(2017)\citenamefont
  {Gon\c{c}alves}, \citenamefont {Capri}, \citenamefont {Rosso}, \citenamefont
  {Ravetti},\ and\ \citenamefont {Atman}}]{goncalves2017quantifying}%
  \BibitemOpen
  \bibfield  {author} {\bibinfo {author} {\bibfnamefont {B.~A.}\ \bibnamefont
  {Gon\c{c}alves}}, \bibinfo {author} {\bibfnamefont {L.}~\bibnamefont
  {Capri}}, \bibinfo {author} {\bibfnamefont {O.~A.}\ \bibnamefont {Rosso}},
  \bibinfo {author} {\bibfnamefont {M.~G.}\ \bibnamefont {Ravetti}}, \ and\
  \bibinfo {author} {\bibfnamefont {A.~P.~F.}\ \bibnamefont {Atman}},\
  }\href@noop {} {\enquote {\bibinfo {title} {Quantifying instabilities in
  financial markets},}\ }\bibinfo {howpublished}
  {https://arxiv.org/abs/1704.05499} (\bibinfo {year} {2017})\BibitemShut
  {NoStop}%
\bibitem [{\citenamefont {Serrano}\ \emph {et~al.}(2009)\citenamefont
  {Serrano}, \citenamefont {Bogu\~{n}\'{a}},\ and\ \citenamefont
  {Vespignani}}]{serrano2009extracting}%
  \BibitemOpen
  \bibfield  {author} {\bibinfo {author} {\bibfnamefont {M.~A.}\ \bibnamefont
  {Serrano}}, \bibinfo {author} {\bibfnamefont {M.}~\bibnamefont
  {Bogu\~{n}\'{a}}}, \ and\ \bibinfo {author} {\bibfnamefont {A.}~\bibnamefont
  {Vespignani}},\ }\href {\doibase 10.1073/pnas.0808904106} {\bibfield
  {journal} {\bibinfo  {journal} {PNAS}\ }\textbf {\bibinfo {volume} {106}},\
  \bibinfo {pages} {6483} (\bibinfo {year} {2009})}\BibitemShut {NoStop}%
\bibitem [{\citenamefont {Tumminello}\ \emph {et~al.}(2011)\citenamefont
  {Tumminello}, \citenamefont {Miccich{\`e}}, \citenamefont {Lillo},
  \citenamefont {Piilo},\ and\ \citenamefont
  {Mantegna}}]{tumminello2011statistically}%
  \BibitemOpen
  \bibfield  {author} {\bibinfo {author} {\bibfnamefont {M.}~\bibnamefont
  {Tumminello}}, \bibinfo {author} {\bibfnamefont {S.}~\bibnamefont
  {Miccich{\`e}}}, \bibinfo {author} {\bibfnamefont {F.}~\bibnamefont {Lillo}},
  \bibinfo {author} {\bibfnamefont {J.}~\bibnamefont {Piilo}}, \ and\ \bibinfo
  {author} {\bibfnamefont {R.~N.}\ \bibnamefont {Mantegna}},\ }\href {\doibase
  10.1371/journal.pone.0017994} {\bibfield  {journal} {\bibinfo  {journal}
  {PLoS ONE}\ }\textbf {\bibinfo {volume} {6(3)}},\ \bibinfo {pages} {e17994}
  (\bibinfo {year} {2011})}\BibitemShut {NoStop}%
\bibitem [{\citenamefont {Orsini}\ \emph {et~al.}(2015)\citenamefont {Orsini},
  \citenamefont {Dankulov}, \citenamefont {Colomer-de Sim\'on}, \citenamefont
  {Jamakovic}, \citenamefont {Mahadevan}, \citenamefont {Vahdat}, \citenamefont
  {Bassler}, \citenamefont {Toroczkai}, \citenamefont {Bogu\~n\'a},
  \citenamefont {Caldarelli}, \citenamefont {Fortunato},\ and\ \citenamefont
  {Krioukov}}]{orsini2015quantifying}%
  \BibitemOpen
  \bibfield  {author} {\bibinfo {author} {\bibfnamefont {C.}~\bibnamefont
  {Orsini}}, \bibinfo {author} {\bibfnamefont {M.~M.}\ \bibnamefont
  {Dankulov}}, \bibinfo {author} {\bibfnamefont {P.}~\bibnamefont {Colomer-de
  Sim\'on}}, \bibinfo {author} {\bibfnamefont {A.}~\bibnamefont {Jamakovic}},
  \bibinfo {author} {\bibfnamefont {P.}~\bibnamefont {Mahadevan}}, \bibinfo
  {author} {\bibfnamefont {A.}~\bibnamefont {Vahdat}}, \bibinfo {author}
  {\bibfnamefont {K.~E.}\ \bibnamefont {Bassler}}, \bibinfo {author}
  {\bibfnamefont {Z.}~\bibnamefont {Toroczkai}}, \bibinfo {author}
  {\bibfnamefont {M.}~\bibnamefont {Bogu\~n\'a}}, \bibinfo {author}
  {\bibfnamefont {G.}~\bibnamefont {Caldarelli}}, \bibinfo {author}
  {\bibfnamefont {S.}~\bibnamefont {Fortunato}}, \ and\ \bibinfo {author}
  {\bibfnamefont {D.}~\bibnamefont {Krioukov}},\ }\href {\doibase
  10.1038/ncomms9627} {\bibfield  {journal} {\bibinfo  {journal} {Nature
  Communications}\ }\textbf {\bibinfo {volume} {6}},\ \bibinfo {pages} {8627}
  (\bibinfo {year} {2015})}\BibitemShut {NoStop}%
\bibitem [{\citenamefont {Gualdi}\ \emph {et~al.}(2016)\citenamefont {Gualdi},
  \citenamefont {Cimini}, \citenamefont {Primicerio}, \citenamefont
  {Di~Clemente},\ and\ \citenamefont {Challet}}]{gualdi2016statistically}%
  \BibitemOpen
  \bibfield  {author} {\bibinfo {author} {\bibfnamefont {S.}~\bibnamefont
  {Gualdi}}, \bibinfo {author} {\bibfnamefont {G.}~\bibnamefont {Cimini}},
  \bibinfo {author} {\bibfnamefont {K.}~\bibnamefont {Primicerio}}, \bibinfo
  {author} {\bibfnamefont {R.}~\bibnamefont {Di~Clemente}}, \ and\ \bibinfo
  {author} {\bibfnamefont {D.}~\bibnamefont {Challet}},\ }\href {\doibase
  10.1038/srep39467} {\bibfield  {journal} {\bibinfo  {journal} {Scientific
  Reports}\ }\textbf {\bibinfo {volume} {6}},\ \bibinfo {pages} {39467}
  (\bibinfo {year} {2016})}\BibitemShut {NoStop}%
\bibitem [{\citenamefont {Saracco}\ \emph {et~al.}(2017)\citenamefont
  {Saracco}, \citenamefont {Straka}, \citenamefont {Clemente}, \citenamefont
  {Gabrielli}, \citenamefont {Caldarelli},\ and\ \citenamefont
  {Squartini}}]{saracco2017inferring}%
  \BibitemOpen
  \bibfield  {author} {\bibinfo {author} {\bibfnamefont {F.}~\bibnamefont
  {Saracco}}, \bibinfo {author} {\bibfnamefont {M.~J.}\ \bibnamefont {Straka}},
  \bibinfo {author} {\bibfnamefont {R.~D.}\ \bibnamefont {Clemente}}, \bibinfo
  {author} {\bibfnamefont {A.}~\bibnamefont {Gabrielli}}, \bibinfo {author}
  {\bibfnamefont {G.}~\bibnamefont {Caldarelli}}, \ and\ \bibinfo {author}
  {\bibfnamefont {T.}~\bibnamefont {Squartini}},\ }\href {\doibase
  10.1088/1367-2630/aa6b38} {\bibfield  {journal} {\bibinfo  {journal} {New
  Journal of Physics}\ }\textbf {\bibinfo {volume} {19}},\ \bibinfo {pages}
  {053022} (\bibinfo {year} {2017})}\BibitemShut {NoStop}%
\bibitem [{\citenamefont {Johansen}\ \emph {et~al.}(2000)\citenamefont
  {Johansen}, \citenamefont {Ledoit},\ and\ \citenamefont
  {Sornette}}]{johansen2000crashes}%
  \BibitemOpen
  \bibfield  {author} {\bibinfo {author} {\bibfnamefont {A.}~\bibnamefont
  {Johansen}}, \bibinfo {author} {\bibfnamefont {O.}~\bibnamefont {Ledoit}}, \
  and\ \bibinfo {author} {\bibfnamefont {D.}~\bibnamefont {Sornette}},\ }\href
  {\doibase https://doi.org/10.1142/S0219024900000115} {\bibfield  {journal}
  {\bibinfo  {journal} {International Journal of Theoretical and Applied
  Finance}\ }\textbf {\bibinfo {volume} {3}},\ \bibinfo {pages} {219} (\bibinfo
  {year} {2000})}\BibitemShut {NoStop}%
\bibitem [{\citenamefont {Sornette}(2003)}]{sornette2003critical}%
  \BibitemOpen
  \bibfield  {author} {\bibinfo {author} {\bibfnamefont {D.}~\bibnamefont
  {Sornette}},\ }\href {\doibase 10.1016/S0370-1573(02)00634-8} {\bibfield
  {journal} {\bibinfo  {journal} {Physics Reports}\ }\textbf {\bibinfo {volume}
  {378}},\ \bibinfo {pages} {1} (\bibinfo {year} {2003})}\BibitemShut {NoStop}%
\bibitem [{\citenamefont {Sornette}\ and\ \citenamefont
  {Zhou}(2006)}]{sornette2006predictability}%
  \BibitemOpen
  \bibfield  {author} {\bibinfo {author} {\bibfnamefont {D.}~\bibnamefont
  {Sornette}}\ and\ \bibinfo {author} {\bibfnamefont {W.-X.}\ \bibnamefont
  {Zhou}},\ }\href {\doibase 10.1016/j.ijforecast.2005.02.004} {\bibfield
  {journal} {\bibinfo  {journal} {International Journal of Forecasting}\
  }\textbf {\bibinfo {volume} {22}},\ \bibinfo {pages} {153} (\bibinfo {year}
  {2006})}\BibitemShut {NoStop}%
\bibitem [{\citenamefont {Yan}\ \emph {et~al.}(2012)\citenamefont {Yan},
  \citenamefont {Woodard},\ and\ \citenamefont {Sornette}}]{yan2012diagnosis}%
  \BibitemOpen
  \bibfield  {author} {\bibinfo {author} {\bibfnamefont {W.}~\bibnamefont
  {Yan}}, \bibinfo {author} {\bibfnamefont {R.}~\bibnamefont {Woodard}}, \ and\
  \bibinfo {author} {\bibfnamefont {D.}~\bibnamefont {Sornette}},\ }\href
  {\doibase 10.1016/j.physa.2011.09.019} {\bibfield  {journal} {\bibinfo
  {journal} {Physica A: Statistical Mechanics and its Applications}\ }\textbf
  {\bibinfo {volume} {391}},\ \bibinfo {pages} {1361} (\bibinfo {year}
  {2012})}\BibitemShut {NoStop}%
\bibitem [{\citenamefont {Johansen}\ and\ \citenamefont
  {Juselius}(1990)}]{johansen1990maximum}%
  \BibitemOpen
  \bibfield  {author} {\bibinfo {author} {\bibfnamefont {S.}~\bibnamefont
  {Johansen}}\ and\ \bibinfo {author} {\bibfnamefont {K.}~\bibnamefont
  {Juselius}},\ }\href {\doibase 10.1111/j.1468-0084.1990.mp52002003.x}
  {\bibfield  {journal} {\bibinfo  {journal} {Oxford Bulletin of Economics and
  statistics}\ }\textbf {\bibinfo {volume} {52}},\ \bibinfo {pages} {169}
  (\bibinfo {year} {1990})}\BibitemShut {NoStop}%
\bibitem [{\citenamefont {Bollerslev}\ and\ \citenamefont
  {Wooldridge}(1992)}]{bollerslev1992quasi}%
  \BibitemOpen
  \bibfield  {author} {\bibinfo {author} {\bibfnamefont {T.}~\bibnamefont
  {Bollerslev}}\ and\ \bibinfo {author} {\bibfnamefont {J.~M.}\ \bibnamefont
  {Wooldridge}},\ }\href {\doibase 10.1080/07474939208800229} {\bibfield
  {journal} {\bibinfo  {journal} {Econometric Reviews}\ }\textbf {\bibinfo
  {volume} {11}},\ \bibinfo {pages} {143} (\bibinfo {year} {1992})}\BibitemShut
  {NoStop}%
\bibitem [{\citenamefont {Ling}\ and\ \citenamefont
  {McAleer}(2002)}]{ling2002stationarity}%
  \BibitemOpen
  \bibfield  {author} {\bibinfo {author} {\bibfnamefont {S.}~\bibnamefont
  {Ling}}\ and\ \bibinfo {author} {\bibfnamefont {M.}~\bibnamefont {McAleer}},\
  }\href {\doibase 10.1016/S0304-4076(01)00090-2} {\bibfield  {journal}
  {\bibinfo  {journal} {Journal of Econometrics}\ }\textbf {\bibinfo {volume}
  {106}},\ \bibinfo {pages} {109} (\bibinfo {year} {2002})}\BibitemShut
  {NoStop}%
\bibitem [{\citenamefont {Meil{\u{a}}}(2003)}]{meila2003comparing}%
  \BibitemOpen
  \bibfield  {author} {\bibinfo {author} {\bibfnamefont {M.}~\bibnamefont
  {Meil{\u{a}}}},\ }\enquote {\bibinfo {title} {Comparing clusterings by the
  variation of information},}\ in\ \href {\doibase
  10.1007/978-3-540-45167-9_14} {\emph {\bibinfo {booktitle} {Learning Theory
  and Kernel Machines}}}\ (\bibinfo  {publisher} {Springer},\ \bibinfo {year}
  {2003})\ pp.\ \bibinfo {pages} {173--187}\BibitemShut {NoStop}%
\bibitem [{\citenamefont {Newman}\ \emph {et~al.}(2001)\citenamefont {Newman},
  \citenamefont {Strogatz},\ and\ \citenamefont {Watts}}]{newman2001random}%
  \BibitemOpen
  \bibfield  {author} {\bibinfo {author} {\bibfnamefont {M.~E.~J.}\
  \bibnamefont {Newman}}, \bibinfo {author} {\bibfnamefont {S.~H.}\
  \bibnamefont {Strogatz}}, \ and\ \bibinfo {author} {\bibfnamefont {D.~J.}\
  \bibnamefont {Watts}},\ }\href {\doibase 10.1103/PhysRevE.64.026118}
  {\bibfield  {journal} {\bibinfo  {journal} {Physical Review E}\ }\textbf
  {\bibinfo {volume} {64}},\ \bibinfo {pages} {026118} (\bibinfo {year}
  {2001})}\BibitemShut {NoStop}%
\bibitem [{\citenamefont {Park}\ and\ \citenamefont
  {Newman}(2004)}]{park2004statistical}%
  \BibitemOpen
  \bibfield  {author} {\bibinfo {author} {\bibfnamefont {J.}~\bibnamefont
  {Park}}\ and\ \bibinfo {author} {\bibfnamefont {M.~E.~J.}\ \bibnamefont
  {Newman}},\ }\href {\doibase 10.1103/PhysRevE.70.066117} {\bibfield
  {journal} {\bibinfo  {journal} {Physical Review E}\ }\textbf {\bibinfo
  {volume} {70}},\ \bibinfo {pages} {66117} (\bibinfo {year}
  {2004})}\BibitemShut {NoStop}%
\bibitem [{\citenamefont {Squartini}\ and\ \citenamefont
  {Garlaschelli}(2011)}]{squartini2011analytical}%
  \BibitemOpen
  \bibfield  {author} {\bibinfo {author} {\bibfnamefont {T.}~\bibnamefont
  {Squartini}}\ and\ \bibinfo {author} {\bibfnamefont {D.}~\bibnamefont
  {Garlaschelli}},\ }\href {\doibase 10.1088/1367-2630/13/8/083001} {\bibfield
  {journal} {\bibinfo  {journal} {New Journal of Physics}\ }\textbf {\bibinfo
  {volume} {13}},\ \bibinfo {pages} {083001} (\bibinfo {year}
  {2011})}\BibitemShut {NoStop}%
\end{thebibliography}
\end{document}